%% file: main.tex
\documentclass[eat,twocolumn]{jmlr}

\usepackage{longtable}
\usepackage{makecell}
\usepackage{graphicx}
\usepackage{siunitx}
\usepackage{float}
\usepackage{booktabs}
\usepackage{amssymb}
\usepackage{amsmath}
 \usepackage{xspace}

\theorembodyfont{\upshape}
\theoremheaderfont{\scshape}
\theorempostheader{:}
\theoremsep{\newline}

\jmlrvolume{}
\firstpageno{1}

\jmlryear{2022}
\jmlrworkshop{Machine Learning for Health (ML4H) 2022}

\makeatletter

\newcommand{\Rmnum}[1]{\expandafter\@slowromancap\romannumeral #1@}
\makeatother

\newcommand{\systemname}{UDAMA}

\title[Turning Silver into Gold]{\textit{Turning Silver into Gold}: Domain Adaptation with Noisy Labels for Wearable Cardio-Respiratory Fitness Prediction}

  \author{\Name{Yu Wu} \Email{yw573@cam.ac.uk}\\
  \Name{Dimitris Spathis} \Email{ds806@cam.ac.uk}\\
  \Name{Hong Jia} \Email{hj359@cam.ac.uk}\\
  \addr Department of Computer Science and Technology, University of Cambridge, UK
  \AND
  \Name{Ignacio Perez-Pozuelo} \Email{ip325@cam.ac.uk}\\
  \Name{Tomas I. Gonzales} \Email{tomas.gonzales@mrc-epid.cam.ac.uk}\\
  \Name{Soren Brage} \Email{soren.brage@mrc-epid.cam.ac.uk}\\
  \Name{Nicholas Wareham} \Email{nick.wareham@mrc-epid.cam.ac.uk}\\
  \addr MRC Epidemiology Unit, School of Clinical Medicine, University of Cambridge, UK\\
  \Name{Cecilia Mascolo} \Email{cm542@cam.ac.uk}\\
  \addr Department of Computer Science and Technology, University of Cambridge, UK
 }

\begin{document}

\maketitle
\input{00_abstract}

\input{01_introduction}
\input{03_methods}
\input{05_new_dataset}

\input{06_new_results}

\input{07_conclusion}

\acks{This work is supported by ERC through Project 833296 (EAR) and by Nokia Bell Labs through a donation.  }


\clearpage

\bibliography{main}
\clearpage
\appendix
\input{appendix}

\end{document}

%% file: 00_abstract.tex
\begin{abstract}
Deep learning models have shown great promise in various healthcare applications. However, most models are developed and validated on small-scale datasets, as collecting high-quality (\textit{gold-standard}) labels for health applications is often costly and time-consuming. As a result, these models may suffer from overfitting and not generalize well to unseen data. At the same time, an extensive amount of data with imprecise labels (\textit{silver-standard}) is starting to be generally available, as collected from inexpensive wearables like accelerometers and electrocardiography sensors.
These currently underutilized datasets and labels can be leveraged to produce more accurate clinical models.
In this work, we propose \textbf{\systemname{}}, a novel model with two key components: \textbf{Unsupervised Domain Adaptation} and \textbf{Multi-discriminator Adversarial training}, which leverage noisy data from source domain (\textit{the silver-standard dataset})
to improve \textit{gold-standard} modeling.
We validate our framework on the challenging task of predicting lab-measured maximal oxygen consumption (VO$_{2}$max), the benchmark metric of cardio-respiratory fitness, using free-living wearable sensor data from two cohort studies as inputs. Our experiments show that the proposed framework achieves the best performance of corr = 0.665 $\pm$ 0.04, paving the way for accurate fitness estimation at scale.
\end{abstract}

%% file: 01_introduction.tex
\section{Introduction}

Deep learning (DL) has been widely applied to many healthcare applications, such as sleep stage classification, stress detection, and fitness prediction~\citep{sleep-detection,  stree-detection,fitness-prediction}. 
Collecting high-quality labels (i.e., gold-standard) for healthcare applications may require extensive efforts and can be particularly time-consuming: as a consequence, most of the existing datasets are small-scale. 
Developing DL models with such small-scale datasets often leads to poor performance and lack of model generalization~\citep{Raschka-2018}. At the same time, large-scale datasets with imprecise labels (i.e., silver-standard datasets) are generally available from inexpensive wearable sensors.
For instance, polysomnography (PSG) sensors are used as the gold-standard measurement for sleep stages monitoring. PSG is mainly utilized in a controlled environment since they it requires dedicated equipment and the involvement of clinicians. As a result, such high-quality (gold-standard) datasets are usually small-scale \citep{psg}. Recently, Electrocardiogram (ECG) sensors embedded in wearable devices have gained momentum due to their affordability and high portability, leading to the generation of less-accurate but large-scale (sliver-standard) datasets \citep{ecg}.

High quality labels are crucial for the development and validation of robust clinical models.
The labels of silver-standard datasets often contain estimation noise due to less accurate collection scheme and,
as they are collected from larger populations, they are characterized by distribution shifts, which makes validation against  gold-standard data difficult.
Therefore, a natural question arises: \textit{Can we leverage noisy large-scale sliver-standard datasets to improve DL model validation on gold-standard datasets?}

Transfer learning (TF) is the natural candidate for this problem ~\citep{Xu20}. However, it is difficult for it to handle the problem where the distribution between two tasks is not similar.
Recently, domain adaptation (DA) has achieved some initial success on solving the problem of the distribution mismatch between the source and target domains~\citep{Patricia-2014}.
Specifically, adversarial-based DAs demonstrated state-of-the-art performance on reducing the difference between data distributions \citep{Tzeng-2017,Hassan-2022}, for example for medical imaging tasks~\citep{Venkataramani-2018}.
However, adversarial-based DAs mainly focus on discriminating the source and target domains as a binary classification task but ignore the fine-grained information that resides within the distribution shifts. To this end, in this paper we introduce a multi-discriminator model to simultaneously learn the coarse-grained and fine-grained domain information under the DA scheme.

In particular, we focus on the challenging health application of cardiorespiratory fitness (CRF) prediction employing maximal oxygen consumption (VO$_{2}$max) as the ground truth. 
CRF is a vital indicator of Cardiovascular disease (CVD)~\citep{predictor}, a leading cause of death globally~\citep{Kaptoge2019}.
Collecting gold-standard VO$_{2}$max labels through maximal exercise tests is difficult because it requires participants to reach exhaustion while on a treadmill, using a face mask with a computerized gas analysis system to monitor ventilation and expired gas fractions, as shown in Figure~\ref{fig:workflow}. This is not practical for particular groups, such as the elderly, for instance. Therefore, an alternative submaximal VO$_{2}$max test~\citep{gonzales2020submaximal} promises to capture fitness levels with lower accuracy. 
Although the gold-standard and silver-standard data capture similar information, they suffer from distribution shift.

In our work, we propose a novel unsupervised domain adaptation framework to improve the gold-standard fitness prediction by leveraging large-scale noisy VO$_{2}$max data.

%% file: 03_methods.tex
\section{Methods}
\begin{figure*}

    \centering
    \includegraphics[width=6 in]{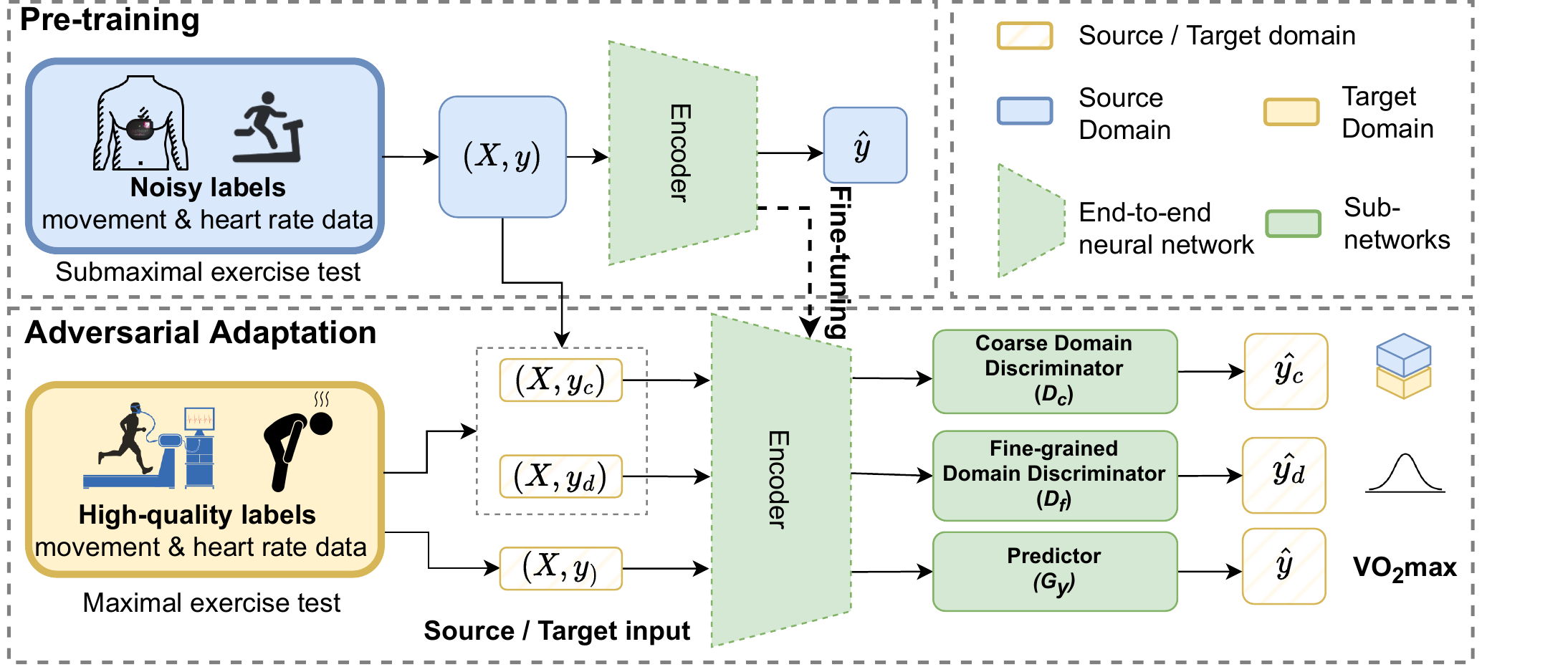}

    \caption{\textbf{Workflow and \systemname{} structure.} }
    \label{fig:workflow}

\end{figure*}

This work introduces a novel unsupervised domain adaptation framework that utilizes multi-discriminator during adversarial training. Herein, the section discussed details of our framework.


The overall model architecture and multi-discriminator training scheme are shown in Figure~\ref{fig:workflow}. In general, our framework consists of two training stages, pre-training on the source domain ${D_s}$ and an adversarial-based training stage on the ${D_t}$ (target domain). 
In the first stage, we build a pre-trained model with noisy health-related labels in a fully supervised setting. Specifically, the neural network is built through multiple stacked bidirectional Gated Recurrent Unit network (GRU) layers to extract time-series features' embedding. After that, Multilayer Perceptron (MLP) layers are followed to extract associated metadata representation.

After pre-training, we fine-tune the large-scale feature embedding on ${D_t}$ after incorporating samples from ${D_s}$. Besides, a novel multiple discriminator-based adversarial domain adaptation learning scheme is trained to distinguish domain labels ${y_c}$ and ${y_d}$ and estimate ${y}$. 
\textbf{Multi-discriminator Adversarial Training.} 
Our framework uses a combination of coarse-grained and fine-grained discriminators to aid the adversarial training for regression tasks. The coarse-grained discriminator (${D_c}$) is similar to other regression DA~\citep{Mathelin-2020,Zhao-2018} and follows the DANN style. In particular, ${D_c}$ mainly focus on discriminating rough domains, which uses a binary domain classifier to discriminate the source of data points (where 0 represents the data comes from the ${D_s}$ and 1 from ${D_t}$). However, the simple binary classes cannot fully represent the data label distribution of each domain. Therefore, we add a more fine-grained discriminator (${D_f}$) to discriminate general distribution differences to aid the adversarial training. 

For adversarial training, the shared encoder first leverages the pre-trained model and extract general feature. Then discriminators are trained to differentiate the domain labels. For ${D_c}$, we optimize ${l_{CSE}}$, a standard cross-entropy loss, to classify the coarse domain source labels ${y_c}$. We can force the extractor to learn a general feature by maximizing such divergence. After that, a ${D_f}$ further maximizes the nuance changes among the sample and updates the discriminator by discriminating the domain distribution label using Gaussian Negative Log-Likelihood (GLL) loss ${l_{GLL}}$~\citep{pytorch}. After that, this multiple two-game adversarial training between the feature encoder and discriminator finally converges into balance. As a result, the feature extractor learns cross-domain knowledge from the multiple discriminators and thus benefits from the main regression prediction task. The whole framework is optimized by the total loss ${D_L}$, which is defined as:
\begin{equation}
L = \alpha l_{MSE} - \lambda_1l_{CSE} -\lambda_2l_{GLL}
\end{equation}
We use ${L}$ to minimize the loss of the predictor while maximizing the loss of domain discriminators. In detail, $\alpha$ is used to scale down the predictor loss to the same level as predictors, $\lambda_1$ and $\lambda_2$ control the relative weight of the discriminator loss, and $\lambda_1$ + $\lambda_2$ = 1. 

%% file: 05_new_dataset.tex
\section{Dataset}\label{experiment}

\textbf{Datasets.} The source domain comes from the silver-standard measurement study (Fenland), including 12,425 participants. It contains a combined heart rate and movement signal from chest ECG sensor Actiheart and noisy VO$_{2}$max labels collected from submaximal exercise tests~\citep{feland}. 
The target domain ${D_t}$ represents the gold-standard measurement dataset BBVS, which is a subset of 191 participants from the Fenland study with directly measured ground-truth VO$_{2}$max~\citep{bvs} during the maximal exercise tests. In the BBVS study, participants need to wear a face mask to measure respiratory gas measurements~\citep{Rietjens-2001} to experience exhaustion tests. 
Further, movement and heart rate are collected using the Actiheart (CamNtech, Papworth, UK) sensor. The details of features and the dataset are in Appendix~\ref{apd:first}

\textbf{Training Strategy.} We evaluate \systemname{} on these two datasets by first pre-training a model on the ${D_s}$ Fenland. 
Second, we develop the adversarial training framework with multi-discriminators on the BVS ${D_t}$. For each domain, feature choices are same as previous study~\citep{Spathis-2022}. 
After the adversarial adaptation, we predict VO$_{2}$max on the held-out test set of BVS using 3-fold cross-validation using \systemname{}. Within each fold, the dataset is split into 70\% training and 30\% testing consisting only of target domain samples.



%% file: 06_new_results.tex
\section{Results and Discussion}\label{6}
\begin{table*}
  \small
  \caption{\textbf{Evaluation of different methods to predict VO$_{2}$max on the BBVS test set.} Each result displays the mean value with standard deviation from three-fold cross-validation. In particular, the results of \systemname{} come from incorporating 5\% amount of BBVS samples From Fenland.}
  \begin{tabular*}{\textwidth}{l @{\extracolsep{\fill}} llll}
  \toprule
    \textbf{Method} & \textbf{R$^2$} & \textbf{Corr} &  \textbf{MSE} & \textbf{MAE} \\ 
    \midrule
    Train from scratch & -0.096 $\pm$ 0.100 & 0.007 $\pm$ 0.250 & 58.048 $\pm$ 10.061  & 6.336 $\pm$ 0.621  \\ 
    Transfer learning (TF)  & 0.283 $\pm$ 0.037  & 0.621 $\pm$ 0.012 & 35.399 $\pm$  5.91 & 4.744 $\pm$ 0.433
 \\ 
    \textbf{UDAMA (ours)} & \textbf{0.392} $\pm$ \textbf{0.07} & \textbf{0.665} $\pm$ \textbf{0.04} &  \textbf{30.794} $\pm$ \textbf{6.805} & \textbf{4.442} $\pm$ \textbf{0.328} \\
    \bottomrule
  \end{tabular*}
  \label{table-results}

\end{table*}


\subsection{VO2max prediction}\label{6.1}
The comparison of the proposed domain adaptation framework and baseline approach for VO$_{2}$max prediction is shown in Table~\ref{table-results}. 
The proposed framework outperforms all baselines for this prediction task and yields the best results among all evaluation metrics. We observe that the correlation (Corr) outperforms the basic transfer learning methods by 7.1\%, the MSE increases by 13.0\%, and R$^2$ improves by 38.5\% compared to the baseline Transfer Learning method. 



\subsection{Domain shift}\label{6.2}
\begin{figure}
    \centering
    \includegraphics[width=3 in]{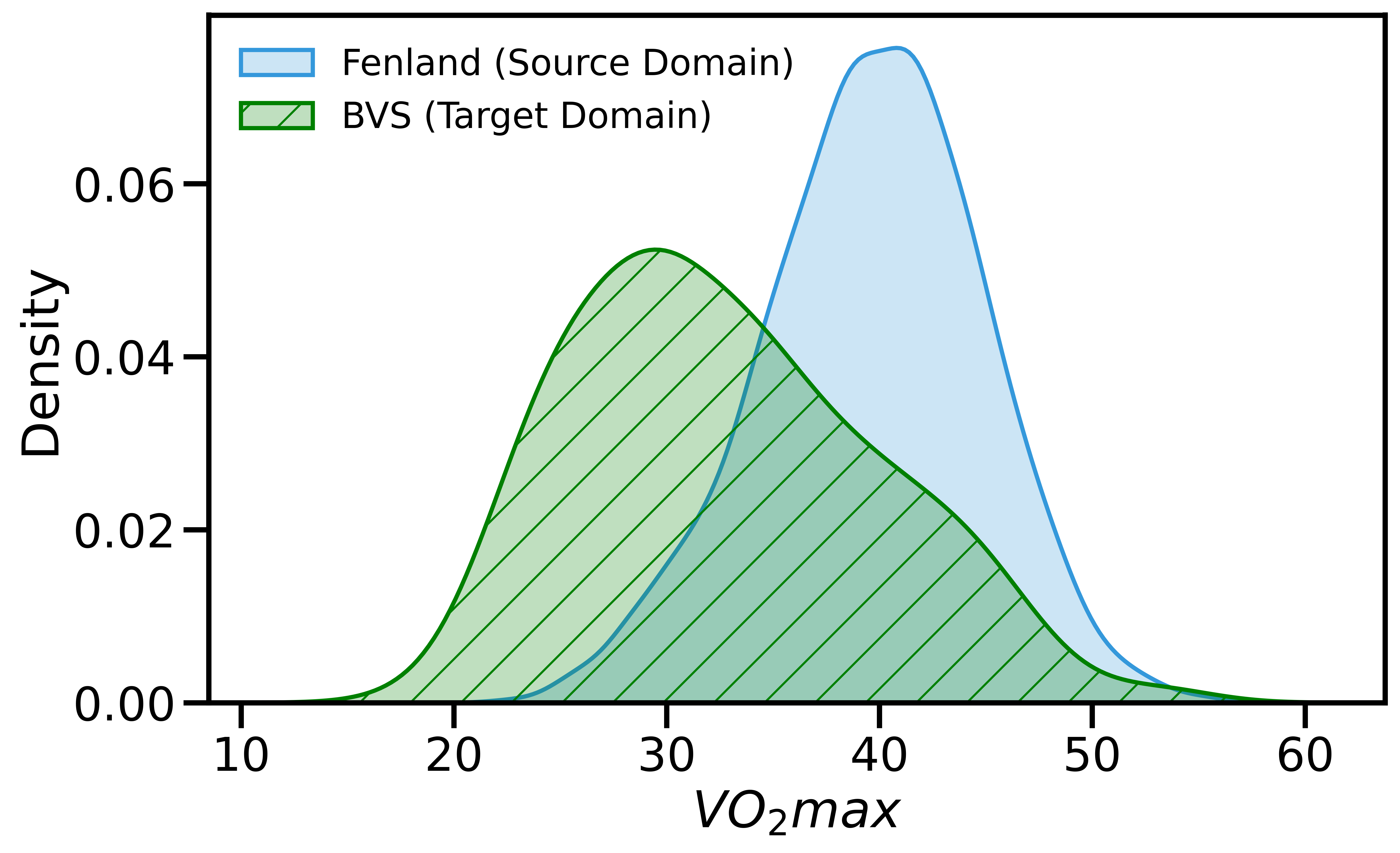}
    \caption{\textbf{Distribution of BBVS and Fenland dataset.}}
    \label{fig:ori-distribution}
\end{figure}

\begin{figure}
    \centering
    \includegraphics[width=3 in]{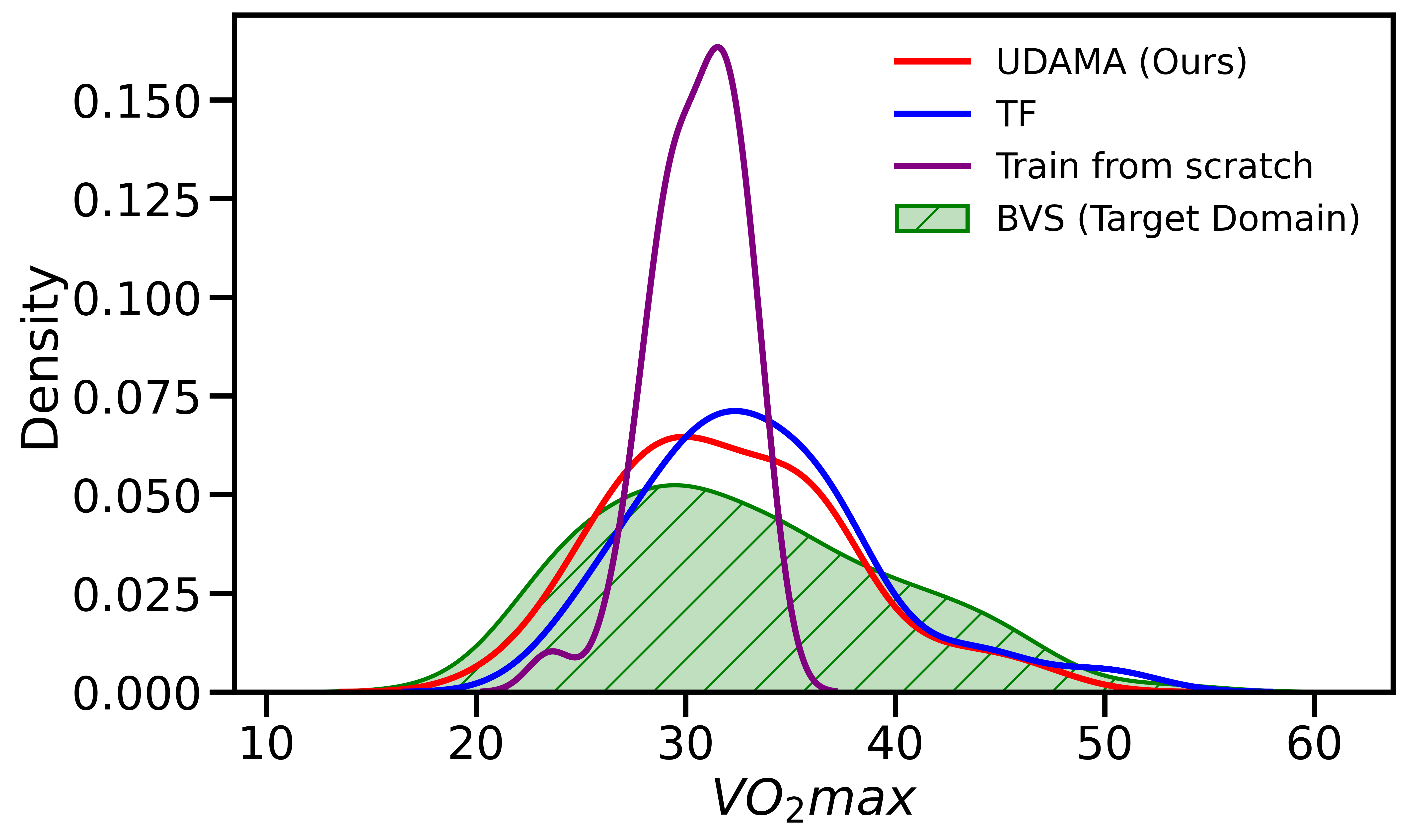}
    \caption{\textbf{Distribution of prediction from different methods and the ground-truth BBVS.}}
    \label{fig:distribution-shift}
\end{figure}

We examined whether \systemname{} can effectively solve the domain shift problem. We observe that our proposed domain adaptation framework can effectively alleviate the problem. Figure~\ref{fig:ori-distribution} shows that the ${D_s}$ dataset (i.e., Fenland) shares a different VO$_{2}$max underlying distribution compared to ${D_t}$ BBVS. Our results in Figure~\ref{fig:distribution-shift} show that our framework learns the small dataset distribution during the adaptation phase and achieves promising results compared with other methods. To examine the distance between domains, we use the Hellinger
Distance (HD) to calculate the similarity of distributions between the prediction and the ground truth. Specifically, our framework's prediction of fitness level lies in the same range as the ground truth, where the mean of the \systemname{} is 33.96\si{\milli\liter\per\kilogram\per\min}, and the mean of BBVS is 32.95\si{\milli\liter\per\kilogram\per\min}.
Besides, distribution ties close compared to baseline methods, where the normalized HD for \systemname{} is 0.188, and HD for TF is 0.264. These results indicate that our framework can effectively alleviate the distribution shift problem of the VO$_{2}$max prediction task. As a result, we reduce the difference between noisy silver-standard data and gold-standard data and improve the gold-standard model performance. 

%% file: 07_conclusion.tex
\section{Conclusion}
In this work, we proposed the unlabeled domain adaptation via multi-discriminator adversarial training framework (\systemname{}) to address the problem of training datasets with various levels of label noise and distribution shifts. By leveraging a large-scale noisy silver-standard dataset and adversarial-based domain adaptation, our proposed method alleviates the domain shit problem and improves the performance on the challenging VO$_{2}$max prediction task. Future work includes generalizing \systemname{} to other healthcare applications, especially regression tasks involving high-dimensional time-series or wearable-sourced data.

%% file: appendix.tex
\section{Appendix}\label{apd:first}
\subsection{Problem Formulation and Notation}\label{3.1}
Here, we denote ${D_s}$ as the source domain containing silver-standard data and ${D_t}$ as the target domain, as shown in Figure~\ref{fig:workflow}. For each domain we assume ${N}$ samples, an input time-series sequence ${X = (x_1,...,x_N) \in \mathbb{R}^{N\times T \times F}}$, the corresponding label is ${y = {(y_c,y_d,y)}}$, where ${y_c}$ represents the coarse-grained binary domain label, ${y_d}$ denotes the fine-grained domain distribution label, and ${y_t}$ denotes the VO$_{2}$max prediction. 
During the adaptation phase, we use multiple discriminators to distinguish ${y_c,y_d}$. In particular, the coarse-grained domain discriminator is ${D_c}$ and the fine-grained domain discriminator is ${D_f}$. 
Also, for the training process, we denote the feature encoder with ${E}$, and the regression predictor with ${G_y}$. 

\subsection{Dataset}
All heart rate data collected during free-living conditions underwent pre-processing for noise filtering~\citep{noise-filter}. If participants had fewer than 72 hours of concurrent wear data (three full days of recording) or inadequate individual calibration data, they were eliminated from the study (treadmill test-based data). Non-wear periods were excluded from the analyses through non-wear detection procedures. This pre-processing algorithm discovered lengthy durations of non-physiological heart rate and extended periods of no movement reported by the device's accelerometer (\textgreater 90 minutes). We used the calculation 1 MET = 71 J/min/kg (3.5 ml O2 min1 kg1) to convert movement intensities into standard metabolic equivalent units (METs). These conversions were then used to classify intensity levels, with behaviors less than 1.5METs classed as sedentary, those between 3 and 6 METs as moderate to vigorous physical activity (MVPA), and those greater than 6METs as vigorous physical activity (VPA). Since time can greatly impact physical activities, we encoded the sensor timestamps using cyclical temporal features \citep{Spathis_self_supervision}. 

Additionally, given the sensors' high sampling rate (1 sample/minute) after matching the HR and Acceleration modalities, learning patterns from such a lengthy sequence (a week's worth of sensor data contains more than 10,000 timesteps) is unfeasible, even with the most potent recurrent neural networks. Therefore, our models downsampled the sampling rate to 15 minutes and used the first 600 timesteps to represent each participant's feature vector. Then each feature vector with 26 features combing time-series and metadata was put into various deep neural networks.